\documentclass[fleqn]{aa54}
\usepackage{graphicx,natbib,txfonts,url}
\bibpunct{(}{)}{;}{a}{}{,}

\newcommand{\figureone}[3]{%
\begin{figure}[tbp]
\begin{center}
\includegraphics[width=88mm]{#1}
\caption{#3}
\label{#2}
\end{center}
\end{figure}
}

\newcommand{\figuretwo}[3]{%
\begin{figure*}[tbp]
\begin{center}
\includegraphics[width=180mm]{#1}
\caption{#3}
\label{#2}
\end{center}
\end{figure*}
}

\def\CaII{\mbox{Ca\,\sc{ii}}} 
\def\HK{\mbox{H\,\&\,K}}
\def\KtwoV{\mbox{K$_{2V}$}}
\def\HtwoV{\mbox{H$_{2V}$}}
\def\CaIIH{\mbox{Ca\,\sc{ii}~H}}

\def\FeI{\mbox{Fe\,\sc{i}}} 

\begin{document}

\title{DOT tomography of the solar atmosphere}
\subtitle{IV. Magnetic patches in internetwork areas}
\titlerunning{Internetwork magnetic patches}
\author{A.G.~de~Wijn\inst{1}
		\and
		R.J~Rutten\inst{1,2}
		\and
		E.M.W.P.~Haverkamp\inst{1}
		\and
		P.~S\"utterlin\inst{1}}
\institute{Sterrekundig Instituut, 
           Utrecht University, 
           Postbus~80\,000, 3508~TA~Utrecht, The~Netherlands\\
           \email{A.G.deWijn@astro.uu.nl, R.J.Rutten@astro.uu.nl}
           \and
           Institute of Theoretical Astrophysics, 
           Oslo University, 
		   P.O.~Box~1029 Blindern, N-0315 Oslo, Norway}
\date{Received 5 May 2005 / Accepted 23 June 2005}
\offprints{A.G.~de~Wijn,\\
           e-mail: {\tt A.G.deWijn@astro.uu.nl}}

\abstract{We use G-band and \CaIIH\ image sequences from the Dutch Open
Telescope (DOT) to study magnetic elements that appear as bright points
in internetwork parts of the quiet solar photosphere and chromosphere.
We find that many of these bright points appear recurrently with varying
intensity and horizontal motion within longer-lived magnetic patches.
We develop an algorithm for detection of the patches and find that all
patches identified last much longer than the granulation.  The patches
outline cell patterns on mesogranular scales, indicating that magnetic
flux tubes are advected by granular flows to mesogranular boundaries.
Statistical analysis of the emergence and disappearance of the patches
points to an average patch lifetime as long as $530\pm50~\mathrm{min}$
(about nine hours), which suggests that the magnetic elements
constituting strong internetwork fields are not generated by a local
turbulent dynamo.  }

\maketitle

\section{Introduction}\label{sec:introduction}

In this paper, we address the appearance and lifetime of magnetic
elements that intermittently show up as bright points in the
internetwork areas of the quiet sun.  The context is the nature of
quiet-sun magnetism, its dynamical coupling to transition-region and
coronal fields, and the existence of a local turbulent dynamo.

The more familiar network bright points are similar.  They constitute
the magnetic network which partially outlines the boundaries of
supergranular cells, and have long been recognized to represent
strong-field magnetic elements, which are traditionally modeled as
flux tubes.  They were first observed as magnetic knots
  \citep{1968SoPh....4..142B} 
and as ``filigree''
  \citep{1973SoPh...33..281D} 
resolved into strings of adjacent bright points by
  \citet{1974SoPh...38...43M}. 
Prompted by 
  \citet{1977SoPh...52..249M}, 
  \citet{1981SoPh...69....9W} 
showed that faculae, filigree, and bright points in wide-band \CaIIH\
filtergrams are manifestations of the same phenomenon.
  \citet{1983SoPh...85..113M}  
subsequently introduced the name ``network bright point'' (NBP) and
initiated an extensive literature observing them as G-band bright
points
  \citep{1984SoPh...94...33M}. 
In particular the G-band studies with the former and the present
Swedish solar telescope on La Palma
  \citep{1995ApJ...454..531B, 
  1998ApJ...495..973B, 
  1998ApJ...506..439B, 
  2004A&A...428..613B, 
  1996ApJ...463..365B, 
  2001ApJ...553..449B, 
  1998ApJ...495..965L, 
  1998ApJ...509..435V, 
  2004A&A...422L..63W, 
  2005A&A...435..327R} 
established that NBPs are brightness manifestations of the small,
strong-field magnetic elements that make up the magnetic network
  \citep{1968SoPh....5..442C, 
  1969SoPh...10..294L, 
  1972SoPh...22..402H, 
  1972SoPh...27..330F, 
  1973SoPh...32...41S}. 
These have been modeled as magnetostatic flux tubes since the pioneering
work of
  \citet{1976SoPh...50..269S, 
  1977PhDT.......237S} 
inspired by
  \citet{1967SoPh....1..478Z}. 
Their hot-wall explanation of the photospheric brightness enhancement
  \citep{1976SoPh...50..269S, 
  1981SoPh...70..207S} 
was recently verified by the MHD simulations of
  \citet{2004ApJ...607L..59K} 
and
  \citet{2004ApJ...610L.137C}, 
which crown a long effort in flux tube modeling
  \citep[e.g.,][]{1988A&A...202..275K, 
  1990A&A...233..583K, 
  1992A&A...262L..29S, 
  1994A&A...285..648G, 
  1998A&A...337..928G, 
  1998ApJ...495..468S, 
  2005A&A...430..691S}. 
NBPs serve as tracers of strong-field flux tubes, especially in the
Fraunhofer G~band (CH lines around 430.5~nm) and the CN band at 388~nm
   \citep[cf.][]{2001ASPC..236..445R}, 
but with the caveat that not all magnetic features produce observable bright
points
  \citep{2001ApJ...553..449B}. 
%

\figuretwo{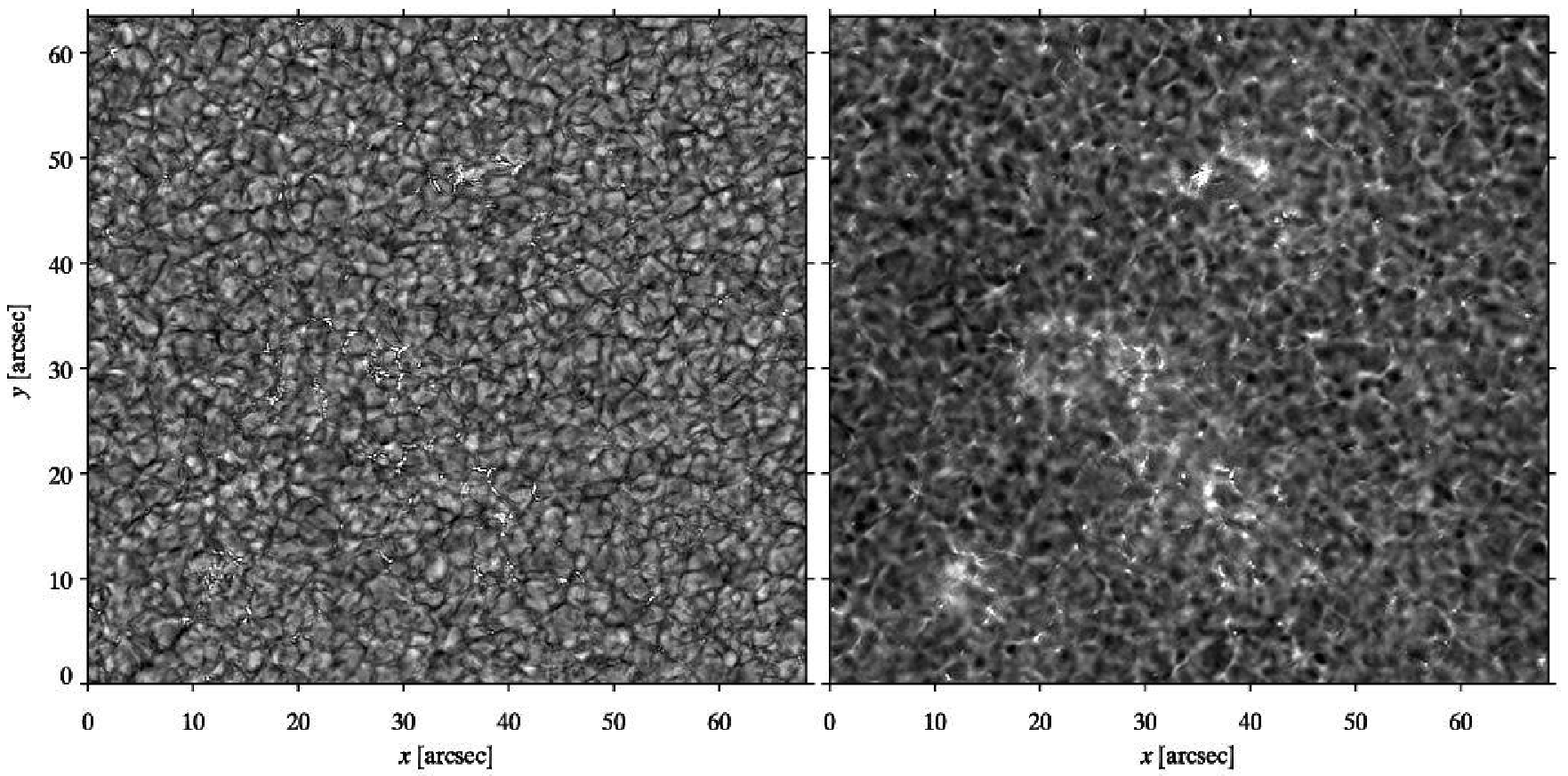}{fig:sample}{A sample image pair from the
Dutch Open Telescope (DOT).  \emph{Left}: G-band image.  \emph{Right}:
co-temporal and co-spatial \CaIIH\ image, clipped to improve contrast.
The network is readily identified in the \CaIIH\ image as regions with
enhanced brightness.  In the G-band image these areas contain many
tiny bright points, often arranged in strings (filigree) within
intergranular lanes. These are NBPs.  Close inspection shows the
presence of isolated IBPs in the internetwork regions.}

The nature of solar magnetism in the quiet-sun internetwork areas, i.e.,
the supergranular cell interiors bordered incompletely by filigree
chains and NBPs that form the network, is less well established, but is
presently under intense scrutiny
  \citep[e.g.,][]{2003ApJ...582L..55D, 
  2003ApJ...597L.177S, 
  2004ApJ...611.1139S, 
  2004ApJ...613..600L, 
  2004ApJ...616..587S, 
  2004Natur.430..326T, 
  2004ApJ...614L..89M}. 
A full spectrum of field strengths seems to be ubiquitously present in
the internetwork at small spatial scales, with the stronger elements
residing in intergranular lanes.  When strong enough, such internetwork
elements are sufficiently evacuated in the low photosphere to appear as
internetwork bright points (IBPs) that are quite similar to NBPs, but
more isolated.  Their existence was already noted by
  \citet{1983SoPh...85..113M}, 
but only recently a more detailed study of photospheric IBPs was presented by
  \citet{2004ApJ...609L..91S}. 
They report a density of 0.3~IBP per Mm$^2$ and lifetimes of a few
minutes, shorter than the average NBP lifetime of 9.3~min measured by
  \citet{1998ApJ...495..973B}. 

It is advantageous to combine photospheric imaging in the G band with
simultaneous co-spatial imaging in the core of the \CaIIH\ line to study
IBP appearance, patterning, and lifetimes.  In \CaII\ \HK\ images
sampling the low chromosphere, both NBPs and IBPs show up with a larger
brightness enhancement over the surrounding area than they do in G-band
or continuum images sampling the photosphere
  \citep[cf.\ Fig.~2 of][]{1999ApJ...517.1013L}. 
\CaIIH\ imaging therefore provides a better diagnostic to detect
isolated IBPs, for which contrast is more important than sharpness.  The
latter is lower for \CaII\ \HK\ images due to strong scattering, and
possibly also due to increasing flux tube width with height.  G-band IBPs
vanish sooner than their \CaII\ counterparts when they diminish in
brightness.  G-band imaging at the highest resolution is needed to
faithfully render the intricate brightness structure of individual
magnetic elements
  \citep{2004A&A...428..613B}, 
but synchronous \CaII\ imaging provides better location detection
especially for isolated ones, as IBPs often are.

Intermittent IBPs observed in \CaII\ \HK\ were called ``persistent
flashers'' by
  \citet{1992ASPC...26..161B, 
  1994ssm..work..251B}, 
who described a particular \CaII\ \KtwoV\ grain which appeared and
disappeared during multiple hours while migrating from the center to the
boundary of a cell.  The grain followed a flow path determined from
independent granulation tracking as if it were a cork floating on the
solar surface.  These flashers figured in the extensive debate whether
all briefly-appearing bright grains in \CaII\ \HK\ image sequences, in
particular at the \HtwoV\ and \KtwoV\ off-center wavelengths, represent
non-magnetic acoustic shocks or magnetism-constrained phenomena
  \citep[e.g.,][]{1982SoPh...80..227S, 
  1991SoPh..134...15R, 
  1995A&A...294..252V, 
  1997ApJ...481..500C, 
  1998SoPh..179..253N, 
  1999ApJ...523..450W, 
  1999ApJ...517.1013L, 
  2000A&A...363..279S}. 
The conclusion is that the acoustic internetwork grains appear only a
few times with a modulation of roughly three minutes, whereas the
strongest internetwork magnetic elements stand out by their ``persistent
flasher'' character: they possess an apparent location memory which
significantly exceeds a few three-minute cycles.  Many such longer-lived
(while modulated) internetwork brightness features were noted in
ultraviolet image sequences from TRACE by
  \citet{2001A&A...379.1052K}, 
who displayed one in their Fig.~7.  A higher-resolution example from
the Dutch Open Telescope (DOT) is shown in Fig.~2 of
  \citet{2004A&A...416..333R}. 
%

\figuretwo{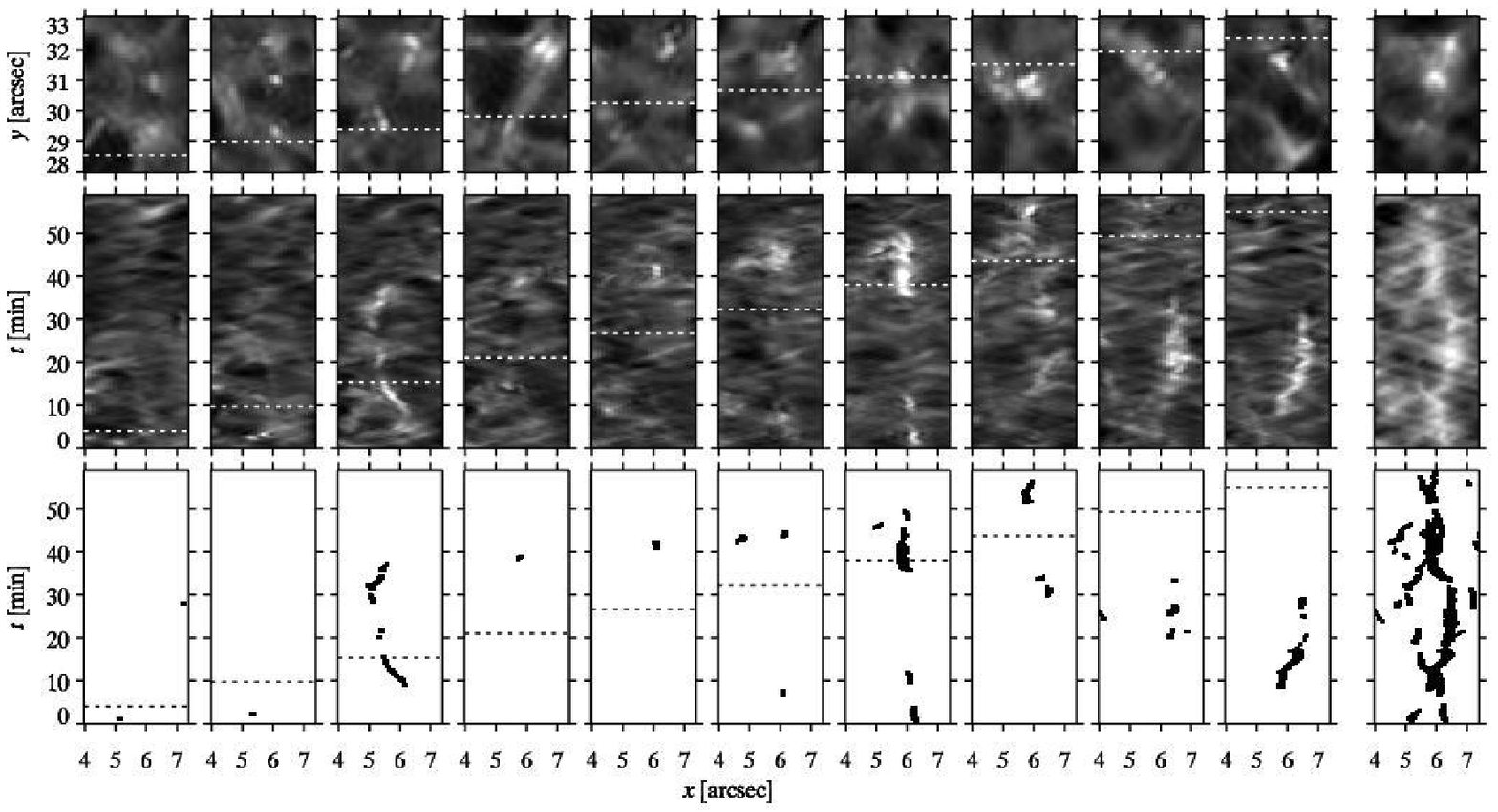}{fig:patchslice}{Example of a magnetic
patch, visualized by partial \CaIIH\ images ($x$--$y$ cutouts, top row),
the associated \CaIIH\ $x$--$t$ slices (middle row), and the
corresponding $x$--$t$ slices from the IBP map sequence (bottom row).
The images shown here are a selection of the data at intervals of
$\Delta t=340~\mathrm{s}$ starting at $t=240~\mathrm{s}$ and $\Delta
y=0.426\arcsec$ starting at $y=0.568\arcsec$, respectively.  The
rightmost panels are averages over all data collected in the duration of
the sequence (top panel) or in the $y$ interval shown here (lower
panels).  The $y$ location of each $x$--$t$ slice is shown by a dashed
line in the associated $x$--$y$ panel, while the time of the latter is
shown by a dashed line in the corresponding $x$--$t$ slice.
}

In this paper, we study IBPs in synchronous G-band and \CaIIH\ image
sequences with high resolution and fast cadence, as a sequel to the
remark by
  \citet{2004A&A...416..333R} 
that the IBP identified as magnetic flasher there persisted over the
full sequence duration of 54~min.  The main issue is whether briefly
appearing IPBs systematically portray longer-lived flux concentrations
that vary temporally in their morphology and intensity.  Our \CaIIH\
imaging and statistical analysis below indeed suggest that this is the
case.  This result is important in the context of field generation by
turbulent dynamos
  \citep{1999ApJ...515L..39C, 
  2001ApJ...560L.197E, 
  2003ApJ...588.1183C}, 
of the field topology surrounding network at chromospheric heights
  \citep[cf.][]{2003ApJ...597L.165S}, 
and of the internetwork contribution to coronal heating by wave
dissipation 
  \citep[e.g.,][]{1983A&A...117..220H} 
and reconnection 
  \citep[e.g.,][]{1988ApJ...330..474P} 
imposed by photospheric foot point motions and topology evolution.


\section{Observations, data reduction, and patch identification}\label{sec:observations}

We use a double image sequence of a quiet area at the disk center
recorded by the DOT from 8:40 to 9:39~UT on June~16, 2003.  The
sequences consist of speckle-reconstructed images taken at a 20-s
cadence in the Fraunhofer G~band with a 10-\AA\ filter centered at
4305~\AA, and synchronous, speckle-reconstructed images taken in the
\CaIIH\ line (3968~\AA) with a narrow-band filter (FWHM 1.3~\AA) at line
center.  A sample pair of G-band and \CaIIH\ images is shown in
Fig.~\ref{fig:sample}.  Details on the telescope, its tomographic
multi-wavelength imaging and image acquisition, and the speckle
reconstruction and standard reduction procedures are given in
  \citet{2004A&A...413.1183R}. 

The images were carefully aligned and destretched using Fourier
correlation.  After clipping to the common field of view
($68\times63~\mathrm{arcsec}^2$), the resulting sequences consist of 178
speckle-reconstructed image pairs of $962\times894$ square pixels of
$0.071\arcsec$ size.  Each sequence was cone-filtered in Fourier space
to remove features that travel with apparent horizontal speed exceeding
the $7~\mathrm{km\,s}^{-1}$ sound speed.  The image sequences may be
downloaded as movies from the DOT
database\footnote{\url{http://dot.astro.uu.nl/}}, together with
synchronous blue and red continuum sequences not used in this analysis.

In order to visually search for recurrent IBPs, we employed an
interactive three-dimensional ``cube slicer'', which dissects both data
cubes simultaneously in $x$--$y$, $x$--$t$, and $y$--$t$ slices with
continuous $(x,y,t)$ selection controlled by mouse movement.  For
example, IBPs show up in the $x$--$t$ slice only when not drifting
in $y$, but a slight wiggling of the $y$ coordinate then helps to track
the three-dimensional ``world-line'' of the feature through both data
cubes.  We so found many IBPs that are intermittently present while
migrating slightly in $x$ and/or $y$.  They are easiest to detect in the
internetwork parts of the \CaIIH\ sequence, but are accompanied, at
least for part of the time, by far sharper co-spatial G-band IBPs in the
underlying photosphere.  Often, we find groups of recurrent IBPs that
appear associated through a shared magnetic structure providing a
longer-term spatial location memory.  We call such linked IBP groups
``magnetic patches''.  They frequently consist of multiple strings of
IBPs that split and merge.

Figure~\ref{fig:patchslice} displays an illustrative example.  The panel
layout mimics cube slicing in the form of a sequence of small successive
$x$--$y$ image cutouts (top row) and a sequence of $x$--$t$ slice
cutouts stepping progressively in their $y$ sampling location (middle
row).  The rightmost column of panels are averages over the plotted $t$
and $y$ ranges.  We discuss two representative cases.  (i)~The first one
is the bright IBP near $y\approx29.5~\mathrm{arcsec}$ in the third
$x$--$y$ panel ($t\approx15~\mathrm{min}$), which is sampled by the
corresponding $x$--$t$ slice.  The latter indicates an IBP lifetime of
about 7~min, during which it migrated leftward at about
$0.15~\mathrm{arcsec\,min}^{-1}$, or $1.8~\mathrm{km\,s}^{-1}$.  It
seems to be fairly isolated, except that some intermittent brightness
appears later at $x=5$--$6~\mathrm{arcsec}$ in the same $x$--$t$ slice.
The next three $x$--$t$ slices show a bright structure near
$t\approx40~\mathrm{min}$ that migrates toward larger $x$.  (ii)~The
second example is the bright grain near $y\approx31~\mathrm{arcsec}$ in
the 7th $x$--$y$ panel for $t\approx38~\mathrm{min}$.  Its $x$--$t$
slice indicates that it appeared abruptly at this time and then lived
for fifteen minutes, but the surrounding slices show enhanced brightness
nearby in $y$ before and after as well.  True cube slicing confirms that
the first example is indeed a continuously present magnetic structure
which merges at $t\approx45~\mathrm{min}$ with the second example.  

The rightmost panels of Fig.~\ref{fig:patchslice} encompass the third
dimension through integration over the full sequence duration (top
panel) or over the $y$ range shown (middle panel).  The nearly
continuous brightness in the interval $x=5$--$7~\mathrm{arcsec}$
suggests that many IBPs in this area belong to a common magnetic patch
throughout the \CaIIH\ sequence duration.  Fig.~\ref{fig:patchsample}
confirms that this is indeed the case with the friends-of-friends patch
definition described below.  Note that the IBPs of cases~(i) and~(ii)
apparently intersect in this panel at $t\approx12~\mathrm{min}$, whereas
they are in fact disjunct in $y$ (as is shown in detail by the $x$--$y$
panels in Fig.~\ref{fig:patchslice} and by the $x$-integrated $y$--$t$
slice in Fig.~\ref{fig:patchsample}).

\figureone{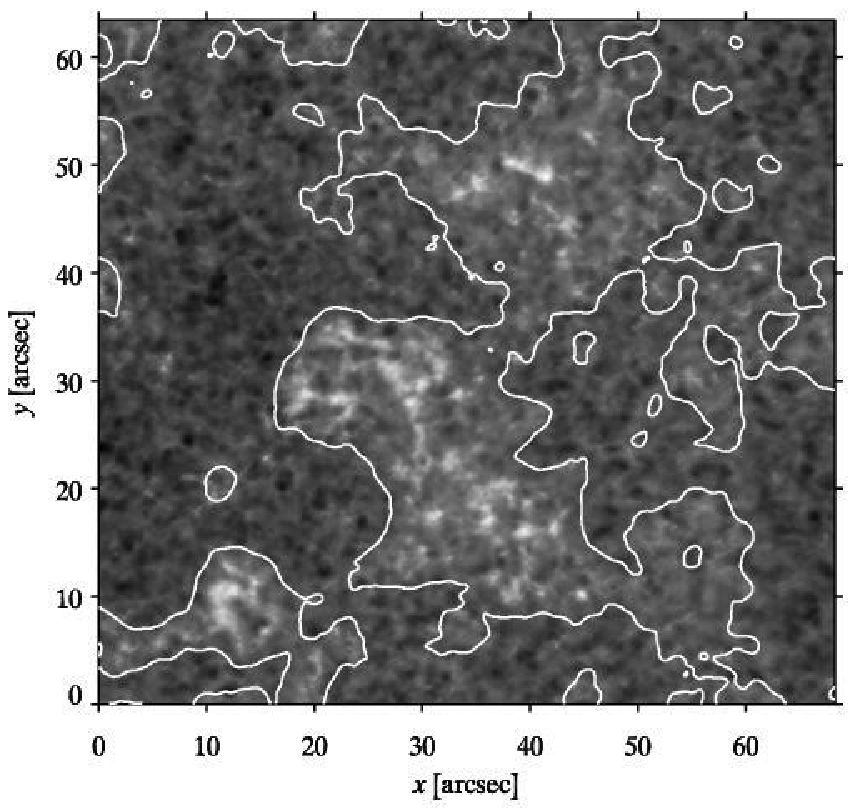}{fig:mask}{The temporal average of the \CaIIH\
image sequence over its one-hour duration with the internetwork mask
contours overlaid in white.  The \CaIIH\ image intensity was scaled
logarithmically in order to show contrast in the internetwork.  The
internetwork mask is computed by taking a 50-pixel boxcar average and
thresholding at the mean value.}

We have developed a detection algorithm to locate magnetic patches made
up by IBPs in order to quantify our visual impressions from cube
slicing.  It includes handling of splits and mergers, such as the
ones discussed above.  First, a mask to block off the network was
constructed from the \CaIIH\ sequence by temporal averaging of the
entire one-hour \CaIIH\ sequence, followed by 50-pixel boxcar smoothing
and thresholding at the mean value, passing only lower values.  The
resulting mask is shown in Fig.~\ref{fig:mask}.  The selection is
conservative in that medium-bright areas around bright network (the
``intermediate'' pixel class of
  \citealp{2001A&A...379.1052K} 
and
  \citealp{2004A&A...416..333R}) 
are also rejected.

We next employed a multi-step procedure to locate IBPs in the masked-off
sequences.  Each image was first convolved with a suitable kernel to
increase the contrast of small round features.  The kernel has the form
\begin{equation}
	K_D(r)=\cos^2\left(\frac{2\pi r}D\right)-\kappa\,,
\end{equation}
where $\kappa$ is a chosen such that the spatial average $\left\langle
K_D(r)\right\rangle=0$.  For the G-band sequence we used a kernel with a
5-pixel diameter ($D=4$).  We chose a 7-pixel kernel ($D=6$) for the
\CaIIH\ images, in which IBPs are larger.  We subsequently produced
binary maps of IBP-candidates by thresholding the convolved images at a
suitable level.  This threshold level must be low, because IBPs
occasionally become very weak between periods of enhanced brightness.
The resulting maps therefore are quite noisy and require further
processing.

The binary map sequences were improved further through spatial and
temporal erosion-dilation processing.  A spatial erosion operation tests
the local nature of a candidate IBP in a binary map by discarding those
pixels whose surroundings do not match a given kernel.  A dilation
operation does the opposite by adding pixels around the candidate IBP
pixels.  For the \CaIIH\ maps, we chose a $3\times3$-pixel kernel for
the erosion as well as the dilation operation, so that only candidate
IBPs of at least this size pass the test.  This spatial processing was
omitted for the G-band maps, because G-band IBPs are often smaller than
this kernel.  However, temporal erosion-dilation processing was done on
both binary-map sequences using a kernel of three time steps
($60~\mathrm{s}$) to remove short-lived features.

Although the erosion-dilation processing removes much noise, there
remain structures in the binary IBP maps that are not IBPs.  Reversed
granulation
  \citep[cf.][]{2004A&A...416..333R} 
in particular produces arc-shaped structures in the binary \CaIIH\ maps
that the above processing fails to remove.  Many of these are
short-lived.  We therefore discarded all structures with lifetimes less
than 80~s (4~images).  

In order to retain only features with a small spatial extent, we keep
only those features whose average maximum instantaneous size in $x$ or
$y$ expressed in units of $0.071\arcsec$ pixels is smaller than the
feature lifetime measured in units of 20-s sampling intervals.  Finally,
we inspected all candidate IBPs visually, either accepting or discarding
them.  We believe that this last, laborious step removes most, if not
all of the misidentifications from our sample.

The third row in Fig.~\ref{fig:patchslice} shows the corresponding
$x$--$t$ slices through the resulting binary \CaIIH\ IBP map sequence.
Much of the brightness pattern in the rightmost slice in the middle row
survives the IBP selection.  The IBPs of case~(i) and (ii) are, of
course, properly detected by our algorithm.  The apparent intersection
at $t\approx12~\mathrm{min}$ also remains.  However, some IBPs
identified by our algorithm, for example those around
$x\approx5~\mathrm{arcsec}$ and $t\approx20~\mathrm{min}$, are not
visible in the $y$-averaged \CaIIH\ brightness.

\figureone{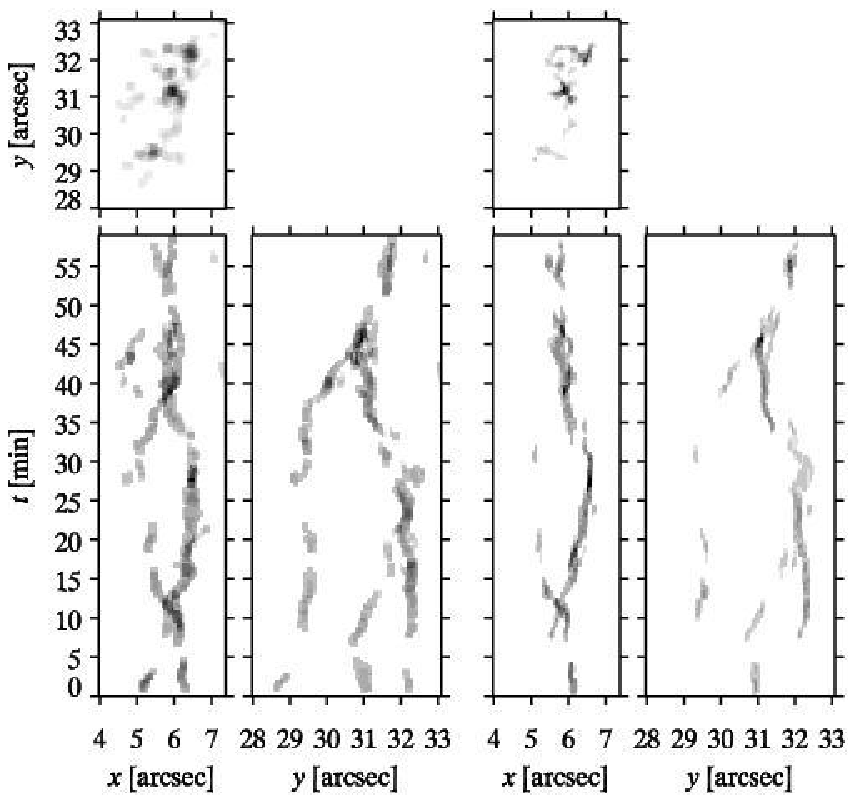}{fig:patchsample}{Integrated $x$--$y$
cutout, and $x$--$t$ and $y$--$t$ slices corresponding to those in
Fig.~\ref{fig:patchslice} of the binary IBP map sequences for the
\CaIIH\ ({\em left}) and G band ({\em right}), showing only those IBPs
that group into the central patch.  Careful comparison with the
rightmost panel in the bottom row of Fig.~\ref{fig:patchslice} shows
that several IBPs present there are part of another patch, such as at
the leftmost edge of the cutout around $t\approx25~\mathrm{min}$.  Each
panel displays the integrated IBP map sequence over the third
coordinate, i.e., the $x$--$y$ map sequence cutouts integrate over time,
the  $x$--$t$ slices over $y$, and the $y$--$t$ slices over $x$.}

The next step is to group the IBPs into magnetic patches.  To this end,
we apply a friends-of-friends algorithm.  Two IBPs are considered
friends if their minimum separation is less than $0.71\arcsec$
(10~pixels), with disregard of temporal seperation.  A patch
then consists of a group of befriended IBPs that have no friends outside
the group.  It may contain IBPs that are not direct friends if a string
of IBPs form a path connecting them.  Each IBP is associated with a
single patch, but a patch often contains multiple IBPs.

The left-hand part of Fig.~\ref{fig:patchsample} shows the result of the
patch analysis for the \CaIIH\ sequence cutout shown in
Fig.~\ref{fig:patchslice}.  The three panels display the corresponding
$x$--$y$ cutout, as well as the associated $x$--$t$ and $y$--$t$ slices
through the binary cube made up of the 24 \CaIIH\ IBPs that are members
of this particular patch.  Each panel is integrated over $x$, $y$, or
$t$, as appropriate, so that a larger blackness signifies the presence
of more IBP pixels in the cube along the third dimension.  Some IBPs
visible in Fig.~\ref{fig:patchslice} are missing, such as the one around
$x\approx4~\mathrm{arcsec}$ and $t\approx25~\mathrm{min}$, because they
are not a member of this patch.  The $y$-integrated $x$--$t$ slice in
the bottom-left panel shows the intersection at
$t\approx12~\mathrm{min}$ and the merger at $t\approx45~\mathrm{min}$
that were already visible in the rightmost panels of
Fig.~\ref{fig:patchslice}.  Adding the information in the $y$--$t$ slice
makes it clear that there are in fact three trails, with mergers at
$t\approx20~\mathrm{min}$ and $t\approx45~\mathrm{min}$.  The apparent
intersection in the $x$--$t$ slice at $t\approx12~\mathrm{min}$ is
actually a disjunct crossing, but the merger at
$t\approx45~\mathrm{min}$ is real.  The latter connects the two legs of
the patch through the friends-of-friends labeling.

Figure~\ref{fig:patchsample} also adds the corresponding triplet for the
G band map sequence (three right-hand panels).  There is a strong
correlation between the two diagnostics, but \CaIIH\ indeed provides
more IBPs.  The right-hand trail is very similar in the two $y$--$t$
slices, but the left-hand trail mostly vanishes in the G-band slice.

\section{Results and Discussion}\label{sec:analysis}

\figureone{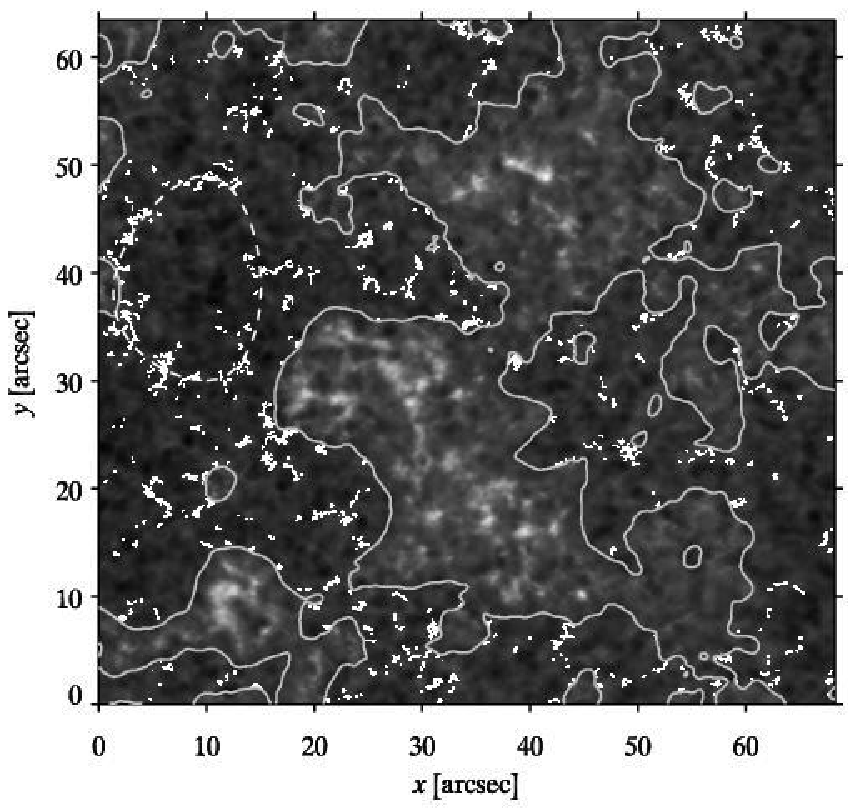}{fig:pmask}{The average \CaIIH\ image with IBPs
identified in the \CaIIH\ sequence overlaid in white.  The IBPs appear
to group in patches that outline edges of cell-like structures, such as
around $(x,y)=(10\arcsec,40\arcsec)$ (indicated by a dashed line).}

\subsection{Patch patterns}

The resulting IBP collection consists of 387 features in the G-band map
sequence that are classified as IPBs, and 848~IBPs in the \CaIIH\ map
sequence.  There are more in the latter because IBPs are more easily
identified in \CaIIH.  We find 149 G-band patches, 76 of which contain
multiple IBPs, and 217 \CaIIH\ patches, 125 of which contain multiple
IBPs.  A comparison shows that 85\% of the G-band patches conincide
spatially with \CaIIH\ patches.

The locations of all \CaIIH\ IBPs, irrespective of their time of
appearance, are shown in Fig.~\ref{fig:pmask}.  They show a striking
pattern in which groups of IBPs appear to partially outline the edges of
cell-like structures of mesogranular scale, e.g., around
$(x,y)=(10\arcsec,40\arcsec)$, where a large, conspicuous cell is marked
with a dashed curve, around $(20\arcsec,15\arcsec)$, and around
$(27\arcsec,42\arcsec)$.

Similar patterns have been reported before in studies of internetwork
magnetograms
  \citep{2003ApJ...582L..55D, 
  2003A&A...411..615S, 
  2003A&A...412L..65D}. 
In particular,
  \citet{2003A&A...412L..65D} 
analyzed a 17-min sequence of \FeI\ magnetograms and found that the
stronger internetwork magnetic elements show a pattern that coincides
with mesogranular upwellings.  One may expect such a pattern to be set
by underlying granular motions.
  \citet{2004A&A...419..757R} 
advected corks by measured granular flows and found that the corks
concentrate at the boundaries of ``trees of fragmenting granules''.
These were previously connected to mesogranules by 
  \citet{2003A&A...409..299R}, 
who found that such trees may have lifetimes of many hours.

\figureone{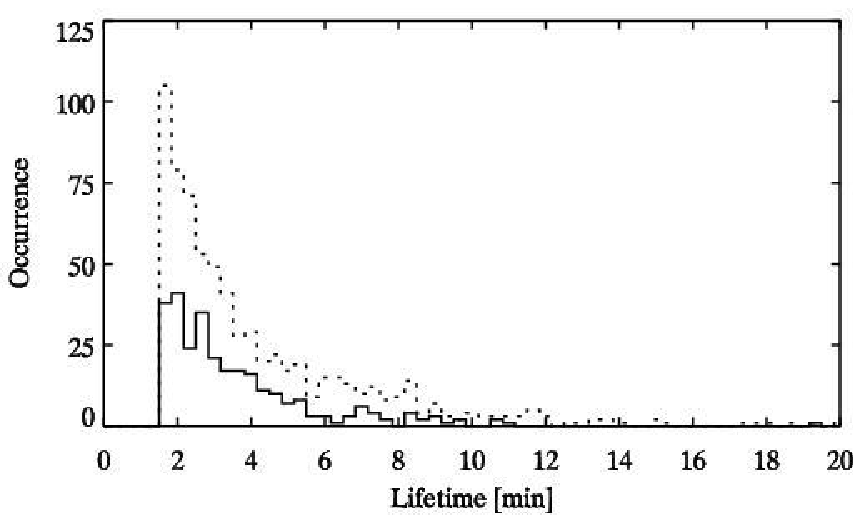}{fig:lthist}{Histograms of IBP lifetimes.
Solid: G-band IBP lifetimes.  Dotted: \CaIIH\ IBP lifetimes.  IBPs with
lifetimes shorter than 4~time steps (80~s) are rejected by our
processing.}

\subsection{IBP lifetimes and number density}

Figure~\ref{fig:lthist} shows a histogram of the durations over which
our algorithm tracks the IBPs that are completely within the temporal
and spatial boundaries, and that do not split or merge.  We find an
average IBP lifetime of $3.5~\mathrm{min}$ in the G band and
$4.3~\mathrm{min}$ in \CaIIH.  Both the G-band and \CaIIH\ IBP lifetime
distributions show a long tail towards long lifetimes, with maxima of
$19.3~\mathrm{min}$ and $25.3~\mathrm{min}$, respectively.

  \citet{1998ApJ...495..973B} 
find longer lifetimes of $9.3~\mathrm{min}$ for G-band network bright
points.  Possibly, their data and analysis permit better tracking of
NBPs.  Their method of NBP detection in G-band images by subtraction of
a suitably scaled continuum image yields a much improved contrast
between granulation and NBPs.  However, there is no equivalent method
for \CaIIH\ images.  In addition, internetwork bright points generally
have substantially lower contrast than network bright points, and are
therefore much harder to identify and track.  By reducing the threshold
value in our data reduction, IBPs mightbe followed through
periods of lower intensity, which somewhat increases the mean lifetime
at the price of many more misidentifications.

Bright point lifetimes in internetwork regions may be shorter than in
network regions.  Where it is strong, the network disturbs the
convection by fragmenting granules into ``abnormal granulation'', so
that flux tubes in network areas have a relatively quiet life compared to
those in internetwork areas, where full-size granules continuously crash
into them.  Their rapid, erratic movements make internetwork IBPs harder
to track, and are furthermore likely to disturb the processes that make
them bright.  Since IBPs cluster in patches, we conclude that, even
though the IBP may momentarily become invisible, the magnetic field
element remains and may become bright again at some later time.  This
agrees well with the conclusion of
  \citet{2001ApJ...553..449B} 
that magnetism is a necessary but not a sufficient condition for the
formation of a network bright point.  We find an average IBP
number density of $0.02~\mathrm{Mm}^{-2}$ in the G-band images and
$0.05~\mathrm{Mm}^{-2}$ in the \CaIIH\ images.  This is an order of
magnitude less than the result of
  \citet{2004ApJ...609L..91S}, 
who find a number density of $0.3~\mathrm{Mm}^{-2}$ in their best G-band
image.  They manually located bright points in data with higher
resolution than used in this analysis, which possibly allowed them to
identify fainter features than our algorithm.  Also, their
$23\arcsec\times35\arcsec$ image contains a small network patch.  We
exclude all network and a fair area around it, where the bright point
number density is increased
  \citep[cf.\ Fig.~2 of][]{2004ApJ...609L..91S}. 

\subsection{Statistical patch lifetime estimation}

Visual inspection of the patches identified by our algorithm shows that
only a few patches begin or end within the duration of the sequence,
indicating that their typical lifetime substantially exceeds one
hour.  We therefore apply statistical arguments to obtain an estimate for
the lifetime.

Assume that we observe a patch with lifetime $\tau$ in a sequence
lasting $t_\mathrm{obs}$.  The patch is visible if the sequence starts
between $t_\mathrm{obs}$ before the patch emerges and the time that the
patch disappears, i.e., in order to see the patch, the sequence must
start in a time window of $\tau+t_\mathrm{obs}$.  We distinguish three
cases.

(i)~A patch persisting throughout the entire sequence.  If the patch has
a lifetime $\tau\ge t_\mathrm{obs}$, it is only visible for the entire
duration of the sequence if it emerges up to $\tau-t_\mathrm{obs}$
before the start of the sequence.  Therefore, the probability of finding
such a patch is given by
\begin{equation}\label{eq:p1}
	p_1(\tau)=\left\{\begin{array}{ll}
		\displaystyle0&\kern1em \mathrm{if}~\tau\le t_\mathrm{obs}\,,\\
		\displaystyle\frac{\tau-t_\mathrm{obs}}{\tau+t_\mathrm{obs}}&
			\kern1em \mathrm{if}~\tau\ge t_\mathrm{obs}\,.
		\end{array}\right.
\end{equation}

(ii)~A patch that emerges as well as disappears within the duration of
the sequence.  A patch with lifetime $\tau\le t_\mathrm{obs}$ emerges as
well as disappears within the duration of the sequence if it emerges up
to $t_\mathrm{obs}-\tau$ after the sequence starts.  The probability
thus is
\begin{equation}\label{eq:p2}
	p_2(\tau)=\left\{\begin{array}{ll}
		\displaystyle\frac{t_\mathrm{obs}-\tau}{\tau+t_\mathrm{obs}}&
			\kern1em \mathrm{if}~\tau\le t_\mathrm{obs}\,,\\[2mm]
		\displaystyle0&\kern1em \mathrm{if}~\tau\ge t_\mathrm{obs}\,.
		\end{array}\right.
\end{equation}

(iii)~A patch that either emerges or disappears within the duration of
the sequence.  In case $\tau\ge t_\mathrm{obs}$, a patch is seen to
emerge, but not to disappear, if the observation is started up to
$t_\mathrm{obs}$ before the patch emerges.  Similarly, it is seen to
disappear, but not to appear, if the observation is started up to
$t_\mathrm{obs}$ before the patch disappears.  In the case that $\tau\le
t_\mathrm{obs}$, the patch is seen to emerge, but not to disappear, if
the patch emerges up to $\tau$ before the end of the observation, and is
seen to disappear, but not to emerge, if the patch disappears up to
$\tau$ after the start of the observation.  The probability $p_3$ of
seeing a patch with lifetime $\tau$ either emerge or disappear in a
sequence of duration $t_\mathrm{obs}$ thus is
\begin{equation}\label{eq:p3}
	p_3(\tau)=\left\{\begin{array}{ll}
		\displaystyle\frac{2\,\tau}{\tau+t_\mathrm{obs}}&
			\kern1em \mathrm{if}~\tau\le t_\mathrm{obs}\,,\\[3mm]
		\displaystyle\frac{2\,t_\mathrm{obs}}{\tau+t_\mathrm{obs}}&
			\kern1em \mathrm{if}~\tau\ge t_\mathrm{obs}\,.
		\end{array}\right.
\end{equation}
We obviously have $p_1(\tau)+p_2(\tau)+p_3(\tau)=1$.

The expected numbers of patches $N'_i$ with $i=1,2,3$ can be computed if
we assume a lifetime distribution.  We adopt a reasonable choice of an
exponential distribution,
\begin{equation}\label{eq:plifedist}
	D_\lambda(\tau)=\lambda\,\mathrm{e}^{-\lambda\tau}\,.
\end{equation}
This distribution has one adjustable parameter, $\lambda$, that is a
measure of the decay time scale of the distribution.  The expected
numbers of patches then follow from integrals of the form
\begin{equation}
	\label{eq:mn}N'_i=
		N\int_0^\infty\!\!p_i(\tau)\,
		D_\lambda(\tau)\,\mathrm{d}\tau\,,
\end{equation}
where $N$ is the total number of patches observed.

To obtain a fit for the parameter $\lambda$, we need to count the number
of patches of each type in our sequence.  In nearly all cases a patch
can be traced much longer by eye than that it is identified by the
algorithm described in Sect.~\ref{sec:observations}.  We therefore
visually inspected the original data and identified those patches that
remain visible during the whole sequence ($N_1=124$), those that both
emerge and disappear during the sequence ($N_2=11$), and those that
either emerge or disappear ($N_3=68$).  We discarded the 14~patches that
enter or exit the field of view during the sequence.  We find an
excellent fit for $\lambda=1.9\times10^{-3}~\mathrm{min}^{-1}$, yielding
$N'_1=123.7$, $N'_2=8.6$, and $N'_3=70.8$.  The excellent fit provides
confidence in the validity of the distribution $D_\lambda(\tau)$.  The
average patch lifetime is then given by
\begin{equation}\label{eq:modellt}
	\langle\tau\rangle=\int_0^\infty\!\!\tau\,D_\lambda(\tau)\,
	\mathrm{d}\tau=\frac1\lambda\approx530\pm50~\mathrm{min}\,,
\end{equation}
where the error was estimated by variation of $N_i$ within their error
bounds set by Poisson statistics.

\section{Conclusion}\label{sec:conclusion}

We have found numerous IBPs in photospheric and chromospheric quiet-sun
internetwork cells.  Our results show that \CaIIH\ line-core filtergrams
are well-suited for finding magnetic flux tubes in the quiet-sun
internetwork.  IBPs in our \CaIIH\ images show a good correlation with
G-band IBPs.  We follow
  \citet{2004ApJ...609L..91S} 
in attributing both IBPs in the G-band and \CaIIH\ sequences to
kiloGauss magnetic flux tubes.

The IBP density that we measure is significantly lower than the value of
  \citet{2004ApJ...609L..91S}, 
which is likely due to our lower resolution and more conservative
identification method.

We find that magnetic IBPs cluster into patches that are not
homogeneously distributed over the internetwork, but rather seem to
outline cell-like structures similar to the magnetic-element
voids found by
  \citet{2003ApJ...582L..55D} 
marking mesogranular upwellings.  This apparent mesogranular
distribution indicates that the magnetic elements which intermittently
appear as IBPs in a patch have a sufficient long lifetime to assemble at
mesogranular vertexes.  The strongest may eventually make it as NBPs to
the supergranular boundaries.  Indeed, all IBP patches identified by our
algorithm exist on time scales much larger than granular time scales.
This result seems to exclude their attribution to a granular dynamo as
described by, e.g.,
  \citet{1999ApJ...515L..39C}. 

Through statistical analysis, we estimate an average lifetime of about
nine~hours.  This estimate depends on a distribution that admittedly
cannot be verified from these observations.  This would require
observations over much longer duration ($>10~\mathrm{hrs}$).  The
Solar-B mission in its sun-synchronous polar orbit will provide the
high-resolution seeing-free long-duration observations that these
analyses require.

\begin{acknowledgements}
The DOT is operated by Utrecht University at the Spanish Observatorio
del Roque de los Muchachos of the Instituto de Astrof{\'{\i}}sica de
Canarias and is presently funded by Utrecht University, the Netherlands
Organisation for Scientific Research NWO, the Netherlands Graduate
School for Astronomy NOVA, and SOZOU.  The DOT efforts are part of the
European Solar Magnetism Network.
\end{acknowledgements}


\end{document}